\documentclass[12pt]{article}

\usepackage{amsfonts,amsmath}
\usepackage[mathscr]{eucal}

\newcommand{\be}{\begin{equation}}
\newcommand{\ee}{\end{equation}}
\newcommand{\bea}{\begin{eqnarray}}
\newcommand{\eea}{\end{eqnarray}}

\begin{document}

\begin{titlepage}
\vspace*{0.5cm}

\renewcommand{\thefootnote}{$\ddagger$}
\begin{center}
{\LARGE\bf  New particle model in extended space-time}

\vspace{0.5cm}

{\LARGE\bf and covariantization of planar Landau dynamics}

\vspace{2cm}

{\large\bf Sergey Fedoruk},${}^{\dagger}$\footnote{\,\,On leave of absence
from V.N.\,Karazin Kharkov National University, Ukraine}\qquad{\large\bf Jerzy Lukierski}${}^{\ast}$

\vspace{1cm}

${}^{\dagger}${\it Bogoliubov  Laboratory of Theoretical Physics, JINR,}\\
{\it Joliot-Curie 6, 141980 Dubna, Moscow region, Russia} \\
\vspace{0.1cm}

{\tt fedoruk@theor.jinr.ru}\\
\vspace{0.5cm}

${}^{\ast}${\it Institute for Theoretical Physics, University of Wroc{\l}aw,}\\
{\it pl.
Maxa Borna 9, 50-204 Wroc{\l}aw, Poland} \\
\vspace{0.1cm}

{\tt lukier@ift.uni.wroc.pl}\\

\vspace{1.5cm}

19.10.2012

\vspace{0.5cm}
\end{center}
\vspace{0.5cm}
\nopagebreak

\begin{abstract}
\noindent We introduce the Maxwell-invariant extension of $D=4$ relativistic free particle model into
ten-dimensional  Maxwell tensorial space. The new model
after first quantization describes in particular Lorentz frame the planar dynamics
providing Landau orbits in the presence of constant magnetic field.
\end{abstract}



\newpage

\end{titlepage}

\setcounter{footnote}{0}

\setcounter{equation}{0}
\section{Introduction}

\quad\, In this paper we shall consider the enlargement of Poincare symmetry
to Maxwell symmetries \cite{Bac,Schr,B,NegroOlmo} and Minkowski space to Maxwell tensorial space
(ten-dimensional in $D=4$). Following the description of reparametrization-invariant $D=4$ relativistic particle model
we will introduce its enlargement to Maxwell tensorial space. We recall  \cite{Schr,NegroOlmo,BG} that the introduction of additional tensorial degrees
of freedom ($x^\mu \to X^M=(x^\mu, y^{\mu\nu})$) has been suggested by the consideration of constant electromagnetic field backgrounds in Minkowski space-time
but the role and physical interpretation of additional coordinates still it is not clarified.

The Lagrangian for scalar relativistic particle interacting with constant electromagnetic (EM) field $f^{(0)}_{\mu\nu}$ described by the potential
$A^{(0)}_\mu=-\frac12\, f^{(0)}_{\mu\nu}x^\nu$ ($F^{(0)}_{\mu\nu}=\partial_\mu A^{(0)}_\nu-\partial_\nu A^{(0)}_\mu=f^{(0)}_{\mu\nu}$)
is the following\footnote{
We use Minkowski metric with positive signature for time component, i.e. $\eta_{\mu\nu}={\rm diag}(+---)$.
}
\begin{equation}\label{Lag-EM-const}
L = -m\sqrt{\dot x^\mu \dot x_\mu} +e A^{(0)}_\mu \dot x^\mu\,.
\end{equation}

The external constant field $f^{(0)}_{\mu\nu}$ breaks the Poincare symmetry, with preserved only the
Bacry-Combe-Richard (BCR) symmetry algebra \cite{Bac} with six generators
$G=M^{\mu\nu}f^{(0)}_{\mu\nu}$, $G=M^{\mu\nu}f^{(0)\,\ast}_{\mu\nu}$, $P_\mu$,
where ($M_{\mu\nu}$, $P_\mu$) describe the Poincare algebra and $f^{(0)\,\ast}_{\mu\nu}=
\frac12\,\epsilon_{\mu\nu\lambda\rho}f^{(0)\,\lambda\rho}$.
The BCR algebra has two-dimensional central extension \cite{ComR} which is integrable to the group
\footnote{Third central extension $A$ defined by $[G,G^\ast]=iA$ can not be lifted to BCR group \cite{Hoog}. }
\begin{equation}\label{CE-BCR}
[ P_\mu,P_\nu]=i\,f^{(0)}_{\mu\nu}\,Z_e+i\,f^{(0)\,\ast}_{\mu\nu}\,Z_g\,,
\end{equation}
where $Z_g\neq 0$ for the particles with nonvanishing magnetic charges and for the model (\ref{Lag-EM-const})
we should put $(Z_e,Z_g)=(e,0)$.

In the formalism with Maxwell symmetries one promotes the constant value of $f^{(0)}_{\mu\nu}$ to new six dynamical degree of freedoms $f_{\mu\nu}=-f_{\nu\mu}$
and introduce corresponding gauge potential with arbitrary values of the space-time independent EM field strength $F_{\mu\nu}=f_{\mu\nu}$
\begin{equation}\label{Pot-EM-const}
A_\mu =-{\textstyle\frac{1}{2}}\, f_{\mu\nu} x^\nu\,.
\end{equation}
If we introduce the covariantized momenta in the presence of such EM field
\begin{equation}
\pi_\mu = p_\mu  +{\textstyle\frac{1}{2}}\,e\, f_{\mu\nu} x^\nu\,,
\end{equation}
where $\{ x^\mu,p_\nu\}_{{}_P}=\delta^\mu_\nu$, we obtain
\begin{equation}
\{ \pi_\mu,\pi_\nu\}_{{}_P}=e\,f_{\mu\nu}\,,
\end{equation}
i.e. after quantization we obtain noncommutative momenta.

It is known \cite{BG,GKL} that the model (\ref{Lag-EM-const}) with $f^{(0)}_{\mu\nu} \to f_{\mu\nu}$
and $A^{(0)}_{\mu} \to A_{\mu}$ (see (\ref{Pot-EM-const}))
provides the realization of the Maxwell algebra, which is obtained from Poincare algebra by the replacement of the
commutative Poincare momentum generators $P_\mu$ by the noncommutative ones
\begin{equation}\label{PP-Z}
[ P_\mu,P_\nu]=ie\,Z_{\mu\nu}\,,\qquad Z_{\mu\nu}=-Z_{\nu\mu}\,,
\end{equation}
where $e$ is the EM coupling constant.

In such a way we pass from BCR algebra which due to central extension (\ref{CE-BCR})
provides by means of local projective
representations the symmetry of quantum-mechanical description of particles in constant EM field, to the Maxwell algebra 
which as well as in the Poincare case does not have a central extension.
It appears that the relation (\ref{CE-BCR}) 
is covariantized, and we obtain the Lorentz-covariant description
in which the particular choices of constant EM field will be
a result of Maxwell and Lorentz symmetry breaking (gauge fixing).
New tensorial generators $Z_{\mu\nu}$ are Abelian and describe so-called
tensorial central charges.
The relations (\ref{PP-Z}) supplemented by the relations
\begin{equation}\label{ZZP-0}
[Z_{\mu\nu},Z_{\lambda\rho}] = [Z_{\mu\nu},P_{\lambda}]=0
\end{equation}
describe the ten-dimensional Lie algebra of Maxwell translations group $\mathcal{G}$.

Let us introduce the group elements of $\mathcal{G}$ by using the exponential parametrization
\begin{equation}\label{M-coset}
G=e^{i\,x^{\,\mu}P_{\mu}}e^{i\,y^{\,\mu\nu}Z_{\mu\nu}}\,.
\end{equation}
and calculate the corresponding Maurer-Cartan one-form
\begin{equation}\label{}
\Omega=-iG^{-1}d G=e^{\,\mu}P_{\mu}+\omega^{\,\mu\nu}Z_{\mu\nu}  \,.
\end{equation}
One gets \footnote{
Always in this paper we use weight coefficient in (anti)symmetrization, i.e.
$A_{(\mu}B_{\nu)}=\frac12\,(A_{\mu}B_{\nu}+A_{\nu}B_{\mu})$, $A_{[\mu}B_{\nu]}=\frac12\,(A_{\mu}B_{\nu}-A_{\nu}B_{\mu})$.
}
\begin{equation}\label{M-om}
e^{\,\mu}= d x^{\mu}\,,\qquad
\omega^{\,\mu\nu}=  d y^{\,\mu\nu} + {\textstyle\frac{1}{2}}\,e\,x^{[\mu}d x^{\nu]} \,.
\end{equation}
The Maxwell-invariant modification of the Lagrangian (\ref{Lag-EM-const}) is introduced by the replacement
\begin{equation}\label{}
A_\mu d x^\mu \qquad \rightarrow \qquad f_{\mu\nu}\omega^{\,\mu\nu}\,,
\end{equation}
i.e. one gets \cite{BG,GKL,GGP}
\begin{equation}\label{Lag-EM-nonconst}
{\cal L} = -m\sqrt{\dot x^\mu \dot x_\mu} +f_{\mu\nu}\left( \dot y^{\,\mu\nu} + {\textstyle\frac{1}{2}}\,e\,x^{[\mu}\dot x^{\nu]} \right)\,.
\end{equation}
The equations of motion following from (\ref{Lag-EM-nonconst}) in the proper time gauge $\dot x^\mu \dot x_\mu=1$ are
\begin{eqnarray}\label{eq-1}
m\ddot x_\mu &=& -f_{\mu\nu}\dot x^\nu\,,
\\[4pt]
\label{eq-2}
\dot y^{\,\mu\nu} &=& - {\textstyle\frac{1}{2}}\,e\,x^{[\mu}\dot x^{\nu]}\,,
\\[4pt]
\label{eq-3}
\dot f_{\mu\nu} &=& 0\,.
\end{eqnarray}
In the realization of the Maxwell algebra described by the model (\ref{Lag-EM-nonconst}) the values
of the generators $Z_{\mu\nu}$ are described by new tensorial variable $f_{\mu\nu}$ \cite{BG,GKL}.

In the simple model  (\ref{Lag-EM-nonconst}) the dynamics in space-time is standard (see (\ref{eq-1})),
with EM interaction described by Lorentz force. The equation (\ref{eq-2}) describes
the motion in tensorial part $y^{\mu\nu}$ of generalized Maxwell space-time $X^M=(x^{\,\mu},y^{\,\mu\nu})$, $M=1,\ldots,10$
and one of the aims of this paper is better understanding of the role of additional tensorial coordinates.
From action  (\ref{Lag-EM-nonconst}) one can derive the constraint
\begin{equation}\label{cons-1}
\phi:=\pi^\mu \pi_\mu -m^2 \approx 0\,,\qquad\qquad \pi_\mu = p_\mu  +{\textstyle\frac{1}{2}}\,e\, f_{\mu\nu} x^\nu\,,
\end{equation}
which describes the numerical value of one of four Maxwell algebra Casimirs generalizing the mass square Casimir for Poincare algebra.

In this paper we shall consider in tensorial space-time $(x^{\,\mu},y^{\,\mu\nu})$
the Maxwell-invariant particle model which provides the fixed numerical values of three
Maxwell Casimirs $C_1$, $C_2$, $C_3$ defining three first class constraints.
In new model we replace the second part of the Lagrangian (\ref{Lag-EM-nonconst}),
with $f_{\,\mu\nu}$ playing the role of Lagrange multipliers, by two terms
depending only on $x^\mu$, $y^{\,\mu\nu}$ and their derivatives.
After first quantization we shall obtain
three wave equations for generalized wave function $\Psi(x^{\,\mu},y^{\,\mu\nu})$ defined on ten-dimensional generalized space-time.
Passing to generalized Maxwell momenta $(p_{\,\mu},f_{\,\mu\nu})$ \footnote{
It appears that the canonical momenta in tensorial sector $y^{\,\mu\nu}$ of Maxwell space-time are
described by the variable $f_{\,\mu\nu}$, representing constant values of EM field strength.}
and introducing fourth eigenvalue equation specifying remaining  fourth Casimir $C_4$, one obtains by special choice
of Lorentz frame the well-known two-dimensional planar dynamics \cite{Landau,JoLip,Lan-orb}
which describes the Landau orbits in constant magnetic field.
First Maxwell Casimir $C_1$ coincides with mass square $m^2$ of the spinless particle in external
constant EM field and two next Casimirs $C_2$ and $C_3$ define various classes of constant EM fields.
In particular for pure magnetic choice of EM field we should assume $C_3=0$. In such a case the fourth Casimir $C_4$
defines discrete energy values of the particle called the Landau levels.

The plan of our paper is the following.
Firstly, in Sect.\,2 we shall define Maxwell algebra, its four Casimirs and consider Maxwell symmetry transformations
of generalized Maxwell space-time.
In Sect.\,3 we introduce the enlargement of standard reparametrization-invariant $D=4$ particle model to the extended
space-time $(x^{\,\mu},y^{\,\mu\nu})$ which after first quantization provides a dynamical realization of Maxwell algebra
with fixed values of three out of four Casimirs.
After passing to the phase space formulation we shall consider briefly the model with fixed values of all four Casimirs.
In order to interprete physically the proposed particle
dynamics we pass to particularly chosen Lorentz frame and show that the basic dynamical equation providing the Maxwell extension
of Klein-Gordon equation describes planar dynamics characterized by Landau orbits in the presence of constant
magnetic field. The last Section contains some outlook, in particular
proposals of the generalization of presented results.

\setcounter{equation}{0}
\section{Maxwell algebra and Maxwell tensorial space-time}

\quad\, The Maxwell algebra $\mathfrak{m}$ \cite{Bac,Schr} \footnote{
Maxwell algebra under different name has been recently rediscovered by other authors (see e.g. \cite{Sor}).
}
is obtained as the following enlargement of the Poincare algebra
with generators ($P_{\mu}$, $M_{\mu\nu}$) by six Abelian tensorial charges
$Z_{\mu\nu}=-Z_{\nu\mu}=(Z_{\mu\nu})^+$
\begin{equation}\label{M-alg}
[P_{\mu},P_{\nu}] = i\,e\, Z_{\mu\nu},
\qquad \quad [Z_{\mu\nu},Z_{\lambda\rho}] = [Z_{\mu\nu},P_{\lambda}]=0 \,,
\end{equation}
\begin{equation}\label{M-alg1}
[M_{\mu\nu},P_{\lambda}] = 2i\,\eta_{\lambda[\mu}\,P_{\nu]}\,,\qquad
\quad [ M_{\mu\nu}, Z_{\lambda\rho}] =
2i\left( \eta_{\lambda[\mu}\, Z_{\nu]\rho}- \eta_{\rho[\mu}\, Z_{\nu]\lambda}\right)\,,
\end{equation}
where $M_{\mu\nu}=-M_{\nu\mu}=(M_{\mu\nu})^+$ describe the Lorentz algebra generators
(see also footnote ${}^2$)
\begin{equation}\label{Lor-alg}
[ M_{\mu\nu}, M_{\lambda\rho}] =
2i\left( \eta_{\lambda[\mu}\, M_{\nu]\rho}- \eta_{\rho[\mu}\, M_{\nu]\lambda}\right).
\end{equation}
The quantity $e$ in  (\ref{M-alg}) provides EM coupling constant. Thus,
the Maxwell algebra $\mathfrak{m}=(P_{\mu},M_{\mu\nu}, Z_{\mu\nu})$ is a semidirect sum
of Lorentz algebra $\mathfrak{l}=(M_{\mu\nu})\cong sl(2,\mathbb{C})$ and subalgebra  $\mathfrak{g}=(P_{\mu}, Z_{\mu\nu})$:
$$
\mathfrak{m}=\mathfrak{l}\subset\!\!\!\!\!\!+\,\mathfrak{g}\,.
$$
We see that the subalgebra $\mathfrak{g}$  (\ref{M-alg}) (see also  (\ref{PP-Z}), (\ref{ZZP-0})) is an ideal and maximal nilpotent subalgebra
of the full Maxwell algebra $\mathfrak{m}$.

The Casimirs of Maxwell algebra are the following \cite{Bac,Schr,NegroOlmo,Sor}
\begin{eqnarray}\label{Max-Cas1}
C_1 &=&P^{\mu}P_{\mu} +e\,Z^{\mu\nu}  M_{\mu\nu} \,,\\[6pt]
\label{Max-Cas2}
C_2 &=&{\textstyle\frac{1}{2}}\,Z^{\mu\nu} Z_{\mu\nu} \,,\\[6pt]
\label{Max-Cas3}
C_3 &=&{\textstyle\frac{1}{2}}\,Z^{\mu\nu} Z^\ast_{\mu\nu}\,,\\[6pt]
\label{Max-Cas4}
C_4 &=& P^{\mu}P^{\nu} Z^{\ast}_{\mu\lambda} Z_{\nu}^{\ast}{}^{\lambda} +
{\textstyle\frac{1}{4}}\,e\,Z^{\mu\nu} Z^\ast_{\mu\nu}\,M^{\lambda\rho} Z^\ast_{\lambda\rho}
\end{eqnarray}
where $Z^\ast_{\mu\nu}=\frac12\,\epsilon_{\mu\nu\lambda\rho} Z^{\lambda\rho}$ is the dual tensor.

Let us introduce the coordinates $X^M$=($x^{\,\mu}$, $y^{\,\mu\nu}$), $y^{\,\mu\nu}{=}{-}y^{\nu\mu}$ which are dual to the Maxwell algebra
generators $P_{\mu}$, $Z_{\mu\nu}$. The generalized space-time parametrized by the coordinates $X^M$=($x^{\,\mu}$, $y^{\,\mu\nu}$) is
described by the group manifold (\ref{M-coset}), and as well
\begin{equation}\label{M-coset1}
G=\frac{\mathfrak{M}}{O(3,1)}=e^{i\,x^{\,\mu}P_{\mu}}e^{i\,y^{\,\mu\nu}Z_{\mu\nu}}\,,
\end{equation}
where $\mathfrak{M}$ is the Maxwell group manifold with included Lorentz group sector. The Maurer-Cartan one-forms
on $G$ are given by the formulae (\ref{M-om}).
The forms (\ref{M-om}) are invariant with respect to the space-time (parameters $a^\mu$) and tensorial translations (parameters $b^{\mu\nu}=-b^{\nu\mu}$)
\begin{equation}\label{tr-M1}
\delta x^\mu=a^\mu\,,\qquad \delta y^{\mu\nu}=b^{\mu\nu}-{\textstyle\frac12}\,e\,a^{[\mu} x^{\nu]}
\end{equation}
and covariant with respect Lorentz transformations (parameters $\ell^{\mu\nu}=-\ell^{\nu\mu}$)
\begin{equation}\label{tr-M2}
\delta x^\mu=\ell^{\mu}{}_{\lambda}x^\lambda\,,\qquad \delta y^{\mu\nu}=2\ell^{[\mu}{}_{\lambda}y^{\nu]\lambda}\,.
\end{equation}
If we denote the group parameters of $\mathfrak{M}$ by $(\Lambda,a,b)\equiv(\Lambda^{\mu}{}_{\nu},a^\mu,b^{\mu\nu})$ where
$\eta^{\nu\rho}\Lambda^{\mu}{}_{\nu}\Lambda^{\lambda}{}_{\rho}=\eta^{\mu\lambda}$,
the group composition law look as follows
\begin{equation}\label{comp-low}
(\Lambda,a,b)\,(\Lambda^{\,\prime},a^{\,\prime},b^{\,\prime})=(\Lambda\Lambda^{\,\prime}, a+
\Lambda a^{\,\prime}, b+\Lambda b^{\,\prime}+e\,a\wedge(\Lambda a^{\,\prime}))\,,
\end{equation}
where $\wedge$ denotes the antisymmetrization with respect to free vector indices.
Note that the  group composition law  (\ref{comp-low}) has important consequence: the composition of two space-time translations
$a$, $a^\prime$ yields the tensorial translation with parameter $a\wedge a^{\,\prime}$.
{}For constant value of $F_{\mu\nu}$ and quantum-mechanical realization
it defines the phase transformation $\exp\{ie\, a_\mu F^{\mu\nu} a^{\,\prime}_\nu\}$ in projective representations of BCR group (see
details in \cite{Schr,B}).
We also recall that similarly as $D{=}4$ Poincare algebra can be obtained
by Wigner-Inon\"{u} contraction from $AdS$ algebra ${\rm O}(3,2)$,
one can get $D{=}4$ Maxwell algebra by nontrivial contraction
(see \cite{Az}) of the Lie algebra
$O(3,1){\otimes}O(3,2)$ \cite{GKL,GGP,Sor,Luk,Durka} not preserving direct sum structure.

\setcounter{equation}{0}
\section{Maxwell-covariant particle model on Maxwell tensorial space-time}

\subsection{Classical theory and the constraint analysis}

\quad\, Using one-forms (\ref{M-om}) we introduce the Maxwell-invariant model as described by the following action
\begin{equation}\label{act-EM-nonconst}
S =\int d\tau L= -m\int\sqrt{\phantom{(}\!e{\cdot} e\phantom{(}\!} -\kappa\int\sqrt{\phantom{(}\!\omega{\cdot}\omega+i\,\omega{\cdot}\omega^\ast\phantom{(}\!}
-\bar\kappa\int\sqrt{\phantom{(}\!\omega{\cdot}\omega-i\,\omega{\cdot}\omega^\ast\phantom{(}\!}\,.
\end{equation}
In (\ref{act-EM-nonconst}) and below we use the notations $e{\cdot} e=e^{\mu} e_{\mu}$, $\omega{\cdot}\omega=\omega^{\mu\nu} \omega_{\mu\nu}$,
$\omega{\cdot}\omega^\ast=\omega^{\mu\nu} \omega^\ast_{\mu\nu}$ and
$\omega^\ast_{\mu\nu}=\frac12\,\epsilon_{\mu\nu\lambda\rho}\,\omega^{\lambda\rho}$. The real constant $m$ and complex constant $\kappa=\kappa_1{+}i\kappa_2$
($\bar\kappa=\kappa_1{-}i\kappa_2$)
parametrize the model. We will see below that these constant parameters define the real values of three Casimirs of Maxwell algebra.

The canonical momenta are defined by
\begin{eqnarray}\label{mom-x}
p_{\mu} &=&\frac{\delta S}{\delta \dot x^{\mu}}=-m\frac{\dot x_{\mu}}{\sqrt{\dot e{\cdot} \dot e}}
-{\textstyle\frac12}\,e\,f_{\mu\nu}x^{\nu}\,,\\[6pt]
\label{mom-z}
f_{\mu\nu} &=&\frac{\delta S}{\delta \dot y^{\mu\nu}}=
-\kappa\frac{\dot \omega_{\mu\nu}+i\,\dot \omega^\ast_{\mu\nu}}{\sqrt{\dot\omega{\cdot}\dot\omega+i\,\dot\omega{\cdot}\dot\omega^\ast}}
-\bar\kappa\frac{\dot \omega_{\mu\nu}-i\,\dot \omega^\ast_{\mu\nu}}{\sqrt{\dot\omega{\cdot}\dot\omega-i\,\dot\omega{\cdot}\dot\omega^\ast}}\,,
\end{eqnarray}
where we use the notation $e^{\mu}=\dot e^{\mu} d\tau$, $\omega^{\mu\nu}=\dot \omega^{\mu\nu} d\tau$
and in  (\ref{mom-x}) we should insert the value of $f_{\,\mu\nu}$ given by formula (\ref{mom-z}).
Euler-Lagrange equations of motion for the model (\ref{act-EM-nonconst}) can be written as follows
\begin{equation}\label{eqs}
\dot p_\mu  -{\textstyle\frac{1}{2}}\,e\, f_{\mu\nu} \dot x^\nu =0\,,\qquad\qquad
\dot f_{\mu\nu}=0\,,
\end{equation}
where $p_\mu$, $f_{\mu\nu}$ is defined by (\ref{mom-x}), (\ref{mom-z}).
First equation (\ref{eqs}) reproduces (\ref{eq-1}) in the proper time gauge, and second
equation (\ref{eqs}) coincides with the equation (\ref{eq-3}) whereas substitution of (\ref{mom-z}) into
second equation (\ref{eqs}) provides field equations of
second order for the coordinates $y^{\mu\nu}(\tau)$.

{}From the definitions (\ref{mom-x}), (\ref{mom-z}) of the momenta we obtain the following first class constraints
\begin{eqnarray}\label{phi1}
\phi_1 &=&\pi{\cdot}\pi-m^2\approx 0\,,\\[6pt]
\label{phi2}
\phi_2 &=&{\textstyle\frac{1}{2}}\,f{\cdot}f-c_2\approx 0\,,\\[6pt]
\label{phi3}
\phi_3 &=&{\textstyle\frac{1}{2}}\,f{\cdot}f^\ast-c_3\approx 0\,,
\end{eqnarray}
where
\begin{equation}\label{pi}
\pi_\mu= p_\mu  +{\textstyle\frac{1}{2}}\,e\, f_{\mu\nu} x^\nu
\end{equation}
and
\begin{equation}\label{c23}
c_2 = \kappa^2+\bar\kappa^2=2(\kappa_1^2-\kappa_2^2)\,,\qquad c_3 = -i(\kappa^2-\bar\kappa^2)=4\kappa_1\kappa_2\,.
\end{equation}
In the constraints analysis we use the equalities $\dot\omega^{\ast\ast}=-\dot\omega$, $\dot\omega^{\ast}{\cdot}\dot\omega^{\ast}=-\dot\omega{\cdot}\dot\omega$
and further $(\dot\omega+i\dot\omega^{\ast}){\cdot}(\dot\omega-i\dot\omega^{\ast})=0$.

The dynamics of our model is fully described
by the constraints (\ref{phi1})-(\ref{phi3}) which appear to be first class.
Indeed, using canonical Poison brackets
\begin{equation}\label{PB}
\{ x^\mu,p_\nu\}_{{}_P}=\delta^\mu_\nu\,, \qquad \{ y^{\,\mu\nu},f_{\lambda\rho}\}_{{}_P}=\delta^{[\mu}_\lambda\delta^{\nu]}_\rho\,,
\end{equation}
one can show that the constraints (\ref{phi1})-(\ref{phi3}) commute between themselves.
Since due to the local reparametrization invariance of the model  (\ref{act-EM-nonconst}) on the world line
the canonical Hamiltonian $H_0{=}p_{\mu}\dot x^{\mu} {+}f_{\mu\nu}\dot y^{\,\mu\nu}{-}L$ is vanishing, $H_0{=}0$,
after quantization the wave function is not depending on proper time coordinate $\tau$.
As in standard case of relativistic particle,
the Klein-Gordon-like constraint (\ref{phi1}) generates local $\tau$-reparamatrization of
space-time coordinates $x_\mu$. The constraints (\ref{phi2})-(\ref{phi3}) yield local transformations only
in tensorial sector which can be represented by $\delta (y+iy^\ast)=\alpha\,(\tau)(f+if^\ast)$ and $\delta (y-iy^\ast)=\bar{ \alpha\,}(\tau)(f-if^\ast)$.

The constraints (\ref{phi1})-(\ref{phi3}) represent eigenvalue equations for first three Casimirs (\ref{Max-Cas1})-(\ref{Max-Cas3}).
The Noether charges generating
the Maxwell transformations (\ref{tr-M1})-(\ref{tr-M2}) in the model are
\begin{eqnarray}
{\cal P}_{\mu} &=& p_\mu  -{\textstyle\frac{1}{2}}\,e\, f_{\mu\nu} x^\nu\,,\label{char-P}\\[4pt]
{\cal M}_{\mu\nu} &=& 2\left(x_{[\mu}p_{\nu]}+2y_{[\mu}{}^{\lambda} f_{\nu]\lambda} \right)\,,\label{char-M}\\[4pt]
{\cal Z}_{\mu\nu} &=& -f_{\mu\nu}\,.\label{char-Z}
\end{eqnarray}
Their Poisson brackets describe classical counterpart of Maxwell  algebra (\ref{M-alg})-(\ref{Lor-alg}),
without imaginary units on the right hand sides.
{}For the realization (\ref{char-P})-(\ref{char-Z}) of Maxwell generators there is valid the following important equality
\begin{equation}\label{}
C_1={\cal P}^{\mu}{\cal P}_{\mu} +e\, {\cal M}^{\mu\nu} {\cal Z}_{\mu\nu}=\pi^\mu \pi_\mu\,.
\end{equation}
Thus, the constraint  (\ref{phi1}) describes the Maxwell phase space realization of first Casimir $C_1$
with constant eigenvalue $m^2$,  whereas remaining constraints  (\ref{phi2}), (\ref{phi3})
realize the Casimirs $C_2$, $C_3$ with eigenvalues (\ref{c23}).
Note also that the equations of motion (\ref{eqs}) are the conservation laws for the Noether charges
($\dot{\cal P}_{\mu}=0$, $\dot{\cal Z}_{\mu\nu}=0$).

Now we shall present the Maxwell transformations of ten-dimensional momentum variables which will be used below.
Taking into account the transformations (\ref{tr-M1}) and the expressions  (\ref{mom-x}), (\ref{mom-z})
we find that they do not change under tensorial translations, generated by ${\cal Z}_{\mu\nu}$, and
\begin{equation}\label{tr-mom}
\delta p_\mu={\textstyle\frac12}\,e\,a^{\nu}f_{\nu\mu}+\ell_{\mu}{}^{\lambda}p_\lambda\,,\qquad\quad \delta f_{\mu\nu}=
2\ell_{[\mu}{}^{\lambda}f_{\nu]\lambda}\,.
\end{equation}
The variables (\ref{pi}) are the covariant momenta. They are invariant with respect to space-time translations (with parameter $a^\mu$),
tensorial translations (with parameter $b^{\mu\nu}$) and transform linearly  with respect Lorentz transformations
\begin{equation}\label{tr-pi}
\delta \pi_\mu=\ell_{\mu}{}^{\lambda}\pi_\lambda\,.
\end{equation}

The model (\ref{act-EM-nonconst}) has the following equivalent first order formulation in phase space
\begin{equation}\label{act-EM-nonconst1}
S = \int\Big(\pi_\mu e^\mu +f_{\mu\nu}\omega^{\mu\nu}\Big) +
{\textstyle\frac12}\int d\tau \Big[g_1\Big(\pi{\cdot}\pi-m^2\Big)+ g_2\Big(f{\cdot}f-2c_2\Big)+g_3\Big(f{\cdot}f^\ast-2c_3\Big)\Big]\,,
\end{equation}
where $\pi_\mu$, $f_{\mu\nu}$ are independent variables and $g_1$, $g_2$ and $g_3$ are
the Lagrange multipliers. Eliminating the variables $\pi_\mu$ and $f_{\mu\nu}$ by their algebraic equations of motion,
and subsequently  expressing on-shell values of $g_1$, $g_2$ and $g_3$
as local functions of $x^\mu$, $y^{\mu\nu}$ and their derivatives, we obtain
the second order action (\ref{act-EM-nonconst}) in Maxwell tensorial space-time.

We add the following comments about the equivalent models (\ref{act-EM-nonconst}) and (\ref{act-EM-nonconst1}):

\smallskip

{\bf i)}\,\,
In the action (\ref{act-EM-nonconst}) last two terms can be generalized as follows
$$
\kappa\, \sqrt{a\,(\omega{\cdot}\omega)+ b\,(\omega{\cdot}\omega^\ast)\!\!\!{\phantom{\bar b}}}
\,+\,\bar\kappa\, \sqrt{\bar a\,(\omega{\cdot}\omega)+\bar b\,(\omega{\cdot}\omega^\ast)}
$$
where the constants $\kappa$, $a$ and $b$ are complex. The complex value of at least one constant $a$, $b$ is necessary, since
the derivation of both constraints (\ref{phi2}) and (\ref{phi3}) requires the condition $a^2 + b^2=0$.
In (\ref{act-EM-nonconst}) the values of the constants $a$ and $b$ ($a^2 + b^2=0$) are specified without loosing generality.

\smallskip

{\bf ii)}\,\,
We can supplement the first-order action (\ref{act-EM-nonconst1}) with an additional term
\begin{equation}\label{act-EM-nonconst4}
S_4 =
{\textstyle\frac12}\int d\tau g_4\Big(\pi^{\mu}\pi^{\nu} f^{\ast}_{\mu\lambda} f_{\nu}^{\ast}{}^{\lambda}-c_4\Big)
\end{equation}
and consider the model with action $\tilde S=S+S_4$.
The term (\ref{act-EM-nonconst4}) fixes the value of Casimir $C_4$ defined by the expression (\ref{Max-Cas4}),
because in the realizations  (\ref{char-P})-(\ref{char-Z}) one gets the following equality
\begin{equation}\label{iden-2}
C_4={\cal P}^{\mu}{\cal P}^{\nu} {\cal Z}^{\ast}_{\mu\lambda} {\cal Z}_{\nu}^{\ast}{}^{\lambda} +
{\textstyle\frac{1}{4}}\,e\,{\cal Z}^{\mu\nu} {\cal Z}^\ast_{\mu\nu}\,{\cal M}^{\lambda\rho} {\cal Z}^\ast_{\lambda\rho}=
\eta^{\lambda\rho}(\pi^{\mu} f^{\ast}_{\mu\lambda})(\pi^{\nu} f_{\nu\rho}^{\ast})
\end{equation}
where also expression (\ref{pi})
and the equality
\begin{equation}\label{eq-contr}
f^{\mu\lambda}f^\ast_{\nu\lambda}={\textstyle\frac{1}{4}}\,(f^{\rho\lambda}f^\ast_{\rho\lambda})\,\delta^\mu_\nu
\end{equation}
were used.

We shall show in subsection\,3.3 that the eigenvalue $c_4$ of fourth Casimir  (\ref{iden-2}) is proportional
to energy square (see (\ref{c4-hE})) and the quantum-mechanical solutions require that the values of  $c_4$
in (\ref{act-EM-nonconst4}) are suitably quantized, i.e. are discrete.
In such a way we obtain the model in Maxwell generalized momentum space $(p{}_{\,\mu},f_{\,\mu\nu})$ with the action  $\tilde S=S+S_4$ describing the particle dynamics
on Maxwell tensorial space-time with fixed value of energy.
We add that we were not able to obtain the term (\ref{act-EM-nonconst4}) from the generalization of the action (\ref{act-EM-nonconst}),
i.e. the second order formulation of the particle model providing all four Casimirs fixed is not known.

\smallskip

{\bf iii)}\,\,
We can put $m=0$ in the action (\ref{act-EM-nonconst}), i.e. obtain only the dynamics in tensorial part of Maxwell space-time. Then
besides eq. (\ref{phi1}) with $m\,{=}\,0$  the additional constraints
\begin{equation}\label{add-constr}
\pi_\mu= p_\mu  +{\textstyle\frac{1}{2}}\,e\, f_{\mu\nu} x^\nu\approx 0
\end{equation}
will appear. Taking into account the Poisson brackets following from  (\ref{PB})
\begin{equation}
\{ \pi_\mu,\pi_\nu\}_{{}_P}=e\,f_{\mu\nu}\,,
\end{equation}
we see that the constraints  (\ref{add-constr}) are all second class if $\det(f_{\mu\nu})\neq 0$,
but in the case $\det(f_{\mu\nu})= 0$ become the mixture of two first and two second class constraint.
Because $\det(f_{\mu\nu})= -\frac{1}{4}\,(f{\cdot}f^\ast)^2$,  in such a model
the value of the constant $c_3$ in the constraint (\ref{phi3}) determines the number of
second class constraints.
In the case of vanishing constant $c_3=0$ the reducible constraints $f^{\ast\,\mu\nu}\pi_{\nu} \approx0$
are first class but then the expression (\ref{iden-2}) indicates that the eigenvalue of forth Casimir $C_4$
should be equal to zero ($c_4=0$).

\subsection{First-quantized theory}

\quad\, In the Maxwell momentum space the quantized phase space variables can have the following
generalized Schr\"{o}dinger realizations:
\begin{equation}\label{op-mom-rep}
\hat x{}^{\,\mu}=i\frac{\partial}{\partial p_{\,\mu}}\,,\qquad \hat p{}_{\,\mu}=p_{\,\mu}\,,\qquad\quad
\hat y{}^{\,\mu\nu}=i\frac{\partial}{\partial f_{\,\mu\nu}}\,,\qquad \hat f{}_{\,\mu\nu}=f_{\,\mu\nu}\,.
\end{equation}

The components of skew-symmetric tensor $f_{\mu\nu}$ can be expressed by the components of three-vectors
of electric field $\vec{\,E}=(E_i)$, $i=1,2,3$ and of magnetic field $\vec{\,H}=(H_i)$ as follows
\begin{equation}\label{f-EH}
f_{0i}=E_i\,, \qquad f_{ij}=-\epsilon_{ijk}H_k\,,\qquad f^\ast_{0i}=H_i\,, \qquad f^\ast_{ij}=\epsilon_{ijk}E_k\,.
\end{equation}
Then, the Lorentz invariants describing the constraints (\ref{phi2}),  (\ref{phi3}) are
\begin{eqnarray}\label{ff-EH}
f^{\mu\nu}f_{\mu\nu}&=&2\left( H_kH_k- E_kE_k\right)=2\left( \vec{\,H}{}^2- \vec{\,E}{}^2\right),
\\
\label{ffast-EH}
f^{\mu\nu}f^\ast_{\mu\nu}&=&-4 E_kH_k=-4 \vec{\,E}\vec{\,H}\,.
\end{eqnarray}
Note that
\begin{equation}\label{det-f}
\det(f_{\mu\nu})=\det(f^\ast_{\mu\nu})={\textstyle\frac{1}{16}}\,(f^{\mu\nu}f^\ast_{\mu\nu})^2=(\vec{\,E}\vec{\,H})^2\,.
\end{equation}
Physical interpretation of the model depend on the choices of values of Casimir operators.
There are four cases which can be considered (we neglect trivial case $f_{\mu\nu}=0$):
\begin{eqnarray}
{\textbf{\textit{I}\!)}} & \quad\det(f_{\mu\nu})\neq0\,; & \label{case-1}\\[5pt]
{\textbf{\textit{I\!I}\!)}} & \quad \det(f_{\mu\nu})=0 & \left[
\begin{array}{lll}
{\textbf{\textit{a}\!)}} & \quad f^{\mu\nu}f_{\mu\nu}>0 &\quad -\quad\mbox{magnetic case},\\[4pt]
{\textbf{\textit{b}\!)}} & \quad f^{\mu\nu}f_{\mu\nu}<0 &\quad -\quad\mbox{electric case},\\[4pt]
{\textbf{\textit{c}\!)}} & \quad f^{\mu\nu}f_{\mu\nu}=0 &\quad -\quad\mbox{radiation case}.
\end{array}
\right. \label{case-2}
\end{eqnarray}
Below we will consider in detail the case {\textbf{\textit{I\!Ia}\!)}} which in particular Lorentz frames leads to the vanishing electric field
and nonvanishing constant magnetic field.

The important role in the case $\det(f)=0$ is played the four-vector
\begin{equation}\label{q-def}
q_\mu\equiv p^{\,\nu}f^\ast_{\nu\mu}\,.
\end{equation}
Taking into account the transformation rules (\ref{tr-mom}) and the relation (\ref{eq-contr})
we obtain that the four vector $q_\mu$ is invariant with respect to space-time translations (parameter $a^\mu$)
and tensorial translations (parameter $b^{\mu\nu}$). Besides it
transforms linearly under the Lorentz transformations
\begin{equation}\label{tr-q}
\delta q_\mu=\ell_{\mu}{}^{\lambda}q_\lambda\,.
\end{equation}
The four-vector  (\ref{q-def}) has the following description in terms of three-vectors
\begin{equation}\label{q-com}
q_\mu=(q_0,q_i)=\left(\vec{\,p}\vec{\,H}\,,\, p_{\,0}\!\vec{\,H}+\vec{\,p}\,{\times}\vec{\,E} \right),
\end{equation}
where $p_\mu=(p_0,\vec{\,p})$. In the magnetic case one can choose $\vec{\,E}=0$ and the direction of magnetic field
along third axis $\vec{\,H}=(H_1,H_2,H_3)=(0,0,\mathrm{H})$. We obtain in such a frame
\begin{equation}\label{q-com-mag}
q_\mu=(q_0;q_1,q_2,q_3)=\mathrm{H}(p_3;0,0,p_0)\,.
\end{equation}

We will look for the quantum spectrum of our model in the magnetic case  {\textbf{\textit{I\!Ia}\!)}}, under the assumption that $c_3=0$. Wave function
defined on Maxwell ten-dimensional momentum space
\begin{equation}\label{WF}
\Psi=\Psi(p_{\mu},f_{\mu\nu})
\end{equation}
satisfies the following constraint equations
\begin{eqnarray}
\left(C_1-c_1\right)\Psi=0&\qquad\Rightarrow\qquad& \pi^\mu\pi_\mu\, \Psi=m^2\,\Psi\,,\label{1eq}\\[6pt]
\left(C_2-c_2\right)\Psi=0&\qquad\Rightarrow\qquad& f^{\mu\nu}f_{\mu\nu}\, \Psi=2c_2\,\Psi\,,\label{2eq}\\[6pt]
\left(C_3-c_3\right)\Psi=0&\qquad\Rightarrow\qquad& f^{\mu\nu}f^\ast_{\mu\nu}\, \Psi=0\,,\label{3eq}\\[6pt]
\left(C_4-c_4\right)\Psi=0&\qquad\Rightarrow\qquad& q^\mu q_\mu\, \Psi=c_4\,\Psi\,,\label{4eq}
\end{eqnarray}
where $\pi_\mu$ take in first-quantized theory the form of differential operators
\begin{equation}\label{pi-q}
\pi_\mu= p_\mu  +{\textstyle\frac{i}{2}}\,e\, f_{\mu\nu} (\partial/\partial  p_\nu)
\end{equation}
and the vector $q_\mu$ is given by (\ref{q-def}).

Let us make comments about the derivation of the equation (\ref{4eq}). Taking into account  (\ref{iden-2}), we obtain for Maxwell algebra realization
 (\ref{char-P})-(\ref{char-Z})
$C_4=\eta^{\lambda\rho}(\pi^{\mu} f^{\ast}_{\mu\lambda})(\pi^{\nu} f_{\nu\rho}^{\ast})$.
Inserting in this expression the quantities  (\ref{pi-q}) and using the identity
(\ref{eq-contr}) we obtain that $C_4=q^{\mu} q_{\mu}$ plus the term proportional to $f^{\mu\nu}f^\ast_{\mu\nu}$,
but such a term is vanishing due to the constraint (\ref{3eq}).

\subsection{Relation with planar dynamics describing Landau orbits}

\quad \, The equations  (\ref{1eq})-(\ref{4eq}) are Maxwell-covariant, but in order to obtain their physical interpretation,
we shall pass to particularly chosen frame
(see also \cite{Schr}).

For that purpose we pass in magnetic case (see (\ref{case-2}a)) from general tensor $f_{\mu\nu}$ to tensor $\tilde f_{\mu\nu}$ with
$\vec{\,E}=0$ and $\vec{\,H}=(H_1,H_2,H_3)=(0,0,\mathrm{H})$ by suitable Lorentz transformations
\begin{equation}\label{f-La1}
f_{\mu\nu}=\tilde\Lambda_{\,\mu}{}^\lambda \tilde\Lambda_{\,\nu}{}^\rho \tilde f_{\lambda\rho}\,.
\end{equation}
The Lorentz transformations matrix $\tilde\Lambda_{\,\mu}{}^\nu$ is the function of $f_{\mu\nu}$ ($\tilde\Lambda=\tilde\Lambda(f)$)
which due to second equation (\ref{eqs}) is on-shell time-independent.
Since we assumed that the only nonvanishing components of $\tilde f_{\mu\nu}$ are $\tilde f_{12}=-\tilde f_{21}=-\mathrm{H}$,
the relation (\ref{f-La1}) can be represented as follows
\begin{equation}\label{f-uv}
f_{\mu\nu}=-\mathrm{H} \,\epsilon_{ab}\,u^{\,a}_{\,\mu} u^{\,b}_{\,\nu}\,,\qquad \epsilon_{12}=-\epsilon_{21}=1\,,
\end{equation}
where the pair of four-vectors $u^{\,a}_{\,\mu}$ is defined as
\begin{equation}\label{uv}
u^{\,a}_{\,\mu}:=\tilde\Lambda_{\,\mu}{}^a\,,\qquad  a=1,2\,.
\end{equation}
Due to the ortogonality property of the Lorentz matrix $\tilde\Lambda$
they satisfy the conditions
\begin{equation}\label{uv-cond}
u^{\,a\mu}u^{\,b}_{\,\mu}=-\delta^{ab}
\end{equation}
and describe five-dimensional coset ${\rm O}(3,1)/{\rm O}(1,1)$.
The relations  (\ref{f-uv}) are additionally invariant under local gauge ${\rm O}(2)$-transformations in $a=(1,2)$ plane
\begin{equation}\label{uv-gauge}
\delta\,u^{\,1}_{\,\mu}=\cos\varphi\,\, u^{\,2}_{\,\mu}\,,\qquad \delta\,u^{\,2}_{\,\mu}=-\sin\varphi\,\, u^{\,1}_{\,\mu}\,,\qquad
\varphi=\varphi(u)
\end{equation}
and consistently the variables $f_{\mu\nu}$ given by formula (\ref{f-uv}) which are restricted by two constraints $f{\cdot}f=2c_2$ and $f{\cdot}f^\ast=0$
contain only four independent degrees of freedom.
Applying the transformations (\ref{f-La1}) in the equations (\ref{2eq}) and (\ref{3eq})
we see that the constant $c_2$ defines the strength of magnetic field
\begin{equation}\label{c3-h}
c_2=\mathrm{H}^2\,,
\end{equation}
which is constant on-shell (see second equation (\ref{eqs})).
After Lorentz transformation $\tilde\Lambda$ we obtain the four-momenta $\tilde p_{\mu}$ defined by
\begin{equation}\label{p-La1}
p_{\,\mu}=\tilde\Lambda_{\,\mu}{}^\nu \tilde p_{\,\nu} \,,\qquad \tilde p_{\,\mu}=
(\tilde p_{\,0};\tilde p_{\,1},\tilde p_{\,2},\tilde p_{\,3})\,,
\end{equation}
and from  (\ref{q-def}) follows that the vector $q_\mu$ has the form  (\ref{q-com-mag}) in `tilded' momentum components, i.e.
\begin{equation}\label{q-com-mag-t}
\tilde q_\mu=\mathrm{H}\,(\tilde p_{\,3};0,0,\tilde p_{\,0})\,.
\end{equation}

Further one can perform next Lorentz transformation in the plane $(0,3)$ which leads
to the vanishing third space component of the four-momentum
\begin{equation}\label{p-La2}
\tilde p_{\,\mu}=\tilde {\tilde \Lambda}_{\,\mu}{}^\nu \tilde{\tilde p}_{\,\nu}\,,\qquad
\tilde{\tilde {p}}_{\,\mu}=(\tilde{\tilde p}_{\,0};\tilde{\tilde p}_{\,1},\tilde{\tilde p}_{\,2},0)\,.
\end{equation}
Using (\ref{q-com-mag}) we obtain that
\begin{equation}\label{q-com-mag-tt}
\tilde{\tilde{\, q}}_\mu=\mathrm{H}\,(0;0,0,\tilde{\tilde p}_{\,0})\,.
\end{equation}
Thus, we see that the constraint (\ref{4eq}) defines the stationary energy values ${\mathscr{E}}\equiv |{\tilde p}_{\,0}|$ in the `doubly-tilded' frame:
\begin{equation}\label{c4-hE}
c_4=\mathrm{H}^2\,{\mathscr{E}}^2\,.
\end{equation}
Note that the transformations $\tilde {\tilde \Lambda}$ leave invariant the tensor $\tilde f_{\mu\nu}$
with only nonvanishing components $\tilde f_{12}=-\tilde f_{21}$.

For the analysis of the remaining equation (\ref{1eq}) let us compare it with the equation
(\ref{4eq}) in the Lorentz frame after $\tilde\Lambda$-transformations
(\ref{f-La1}).
Subtracting rescaled equation (\ref{4eq}) written in the `tilded' Lorentz frame (\ref{q-com-mag-t})
\begin{equation}\label{eq-4L}
\left[\left(\tilde p_{\,0}\right)^2 -
\left(\tilde p_{\,3}\right)^2\right]\Psi=
{\mathscr{E}}^2\,\Psi
\end{equation}
from the equation  (\ref{1eq})
\begin{equation}\label{eq-1L}
\left[\left(\tilde p_{\,0}\right)^2 -
\left(\tilde p_{\,3}\right)^2-\left(\tilde p_{\,1}  -\frac{ie\mathrm{H}}{2}\,\frac{\partial}{\partial  \tilde p_{\,2}}\right)^2 -
\left(\tilde p_{\,2}  +\frac{ie\mathrm{H}}{2}\,\frac{\partial}{\partial  \tilde p_{\,1}}\right)^2\right]\Psi=
m^2\,\Psi\,,
\end{equation}
we get the two-dimensional equation
\begin{equation}\label{eq-2L}
\left[\left(\tilde p_{\,1}  -\frac{ie\mathrm{H}}{2}\,\frac{\partial}{\partial  \tilde p_{\,2}}\right)^2 +
\left(\tilde p_{\,2}  +\frac{ie\mathrm{H}}{2}\,\frac{\partial}{\partial  \tilde p_{\,1}}\right)^2\right]\Psi=
\varepsilon^{2}\,\Psi\,,\qquad \varepsilon^{2}={\mathscr{E}}^2-m^2\,.
\end{equation}
The equation  (\ref{eq-2L}) is one which describes the two-dimensional planar quantum states
in Landau problem \cite{Landau} if it is derived from Klein-Gordon equation in constant magnetic field \cite{JoLip}.
Let us observe that the operator on left-hand side of (\ref{eq-2L})
can be represented as one-dimensional
oscillator with frequency $\omega=2e\mathrm{H}$ with quantized energy eigenvalues.
If the wave function in  (\ref{eq-2L}) is square-integrable, only discrete values of $\varepsilon^2$
are allowed, which provide the discrete energy levels ${\mathscr{E}}_n$, where
$\varepsilon^2={\mathscr{E}}_n^2-m^2$ is given by $\omega(n+\frac12)=e\mathrm{H}(2n+1)$, $n\in \mathbb{Z}$.
In such a way we obtain quantized energy levels in Landau problem
describing so-called  Landau orbits
\begin{equation}\label{E-n}
{\mathscr{E}}_n=\pm \left[m^2+ e\,\mathrm{H}\,(2n+1) \right]^{1/2}\,, \qquad n\in \mathbb{Z}\,,
\end{equation}
where the two-dimensional relativistic energy-momentum dispersion relation (${\mathscr{E}}=\pm \left(\varepsilon^2+m^2\right)^{1/2}$)
follows from the relativistic Klein-Gordon form of equation (\ref{eq-1L}).

We see that the nontrivial dynamics is confined to $(\tilde p_{\,1},\tilde p_{\,2})$ plane because
in the double-tilded Lorentz frame (see (\ref{p-La2})) we can eliminate the motion along third axis $\tilde{\tilde x}_{\,3}$.
The time-depended wave function in momentum representation is given by the stationary wave functions in $\tilde\Lambda$-frame
which solve equation  (\ref{eq-4L}) with $\tilde{\tilde p}_3=0$
\begin{equation}\label{WF-mom}
\Psi(\tilde p_{\,i},u^{\,a}_{\,\mu};t)=\int d\tilde{\tilde p}_{\,0}\,\delta\Big( \left(\tilde{\tilde p}_{\,0}\right)^2 -
{\mathscr{E}}^2\Big)e^{i\tilde{\tilde p}_{\,0}t}\,\Psi(\tilde p_{\,1},\tilde p_{\,2},u^{\,a}_{\,\mu})\,.
\end{equation}
If we recall that one should select square integrable solutions of  (\ref{eq-2L}), the Dirac delta-function is nonvanishing
in (\ref{WF-mom}) only if ${\mathscr{E}}={\mathscr{E}}_n$, so using the expansion on the Landau states
with discrete energies ${\mathscr{E}}_n$ one gets
\begin{equation}\label{wf-mom-Ls}
\Psi(\tilde p_{\,i},u^{\,a}_{\,\mu};t)=\sum_{n}\psi_n(\tilde p_{\,1},\tilde p_{\,2},u^{\,a}_{\,\mu})\,e^{i{\mathscr{E}}_n t}\,,
\end{equation}
where ${\mathscr{E}}=\pm \tilde{\tilde p}_{\,0}$.

Two vectors $u^{\,a}_{\,\mu}$ define purely kinematical five degrees of freedom describing the Lorentz frame
(see the constraints  (\ref{uv-cond})).\footnote{
The vector variables  $u^{\,a}_{\,\mu}$ describe a selected set of so-called Lorentz vectorial harmonics \cite{Sok,DelGS}.
}
Because the variables $u^{\,a}_{\,\mu}$ are purely kinematical
($u^{\,a}_{\,\mu}\equiv u^{\,a}_{\,\mu}(f)$ due to eq.\,(\ref{eqs}) imply $\dot u^{\,a}_{\,\mu}=0$)
and describe the choice of Lorentz
frame,
we can consider them as constant Lorentz group parameters.

The time-independent norm of the wave function (\ref{wf-mom-Ls}) at fixed time $t$ is given by
\begin{equation}\label{WF-norm-L}
\parallel\!\Psi\!\!\parallel^{\,2} =\sum_n N_n\int \mu(u)\,d^{\,2} \!\tilde p\,
\left|\,\psi_n(\tilde p_{\,1},\tilde p_{\,2},u^{\,a}_{\,\mu})\right|^{\,2} \,,
\end{equation}
where $\mu(u)$ provides invariant measure on the five-dimensional coset ${\rm O}(3,1)/{\rm O}(1,1)$
and $N_n$ describe energy level-dependent normalizations.
The measure defining the norm (\ref{WF-norm-L}) can be described also  in fully Maxwell-covariant way.
For such purpose one should replace the product of measures $d\tilde p_{\,1}\,d\tilde p_{\,2}\,\mu(u)$ for $n$-th Landau level
by the following Maxwell-covariant measure
\begin{equation}\label{meas-cov}
d\tilde p_{\,1}\,d\tilde p_{\,2}\,\mu(u)\,\,\,\sim\,\,\, \delta(f{\cdot}f^\ast)\,\delta(f{\cdot}f-2\mathrm{H}^2)\,
\delta[(p{\cdot}f^\ast)^2-\mathrm{H}^2(m^2+\mathrm{H}(2n+1))] \,d^6 f d^4 p\,,
\end{equation}
where we parametrize $f_{\,\mu\nu}$ in terms of variables $(\mathrm{H},u^{\,a}_{\,\mu})$ (see (\ref{f-uv})).
We obtain the following Maxwell-covariant norm of the wave function $\Psi^{(cov)}(p_{\,\mu},f_{\,\mu\nu})$
in arbitrary Maxwell symmetry frame (see also \cite{Schr,NegroOlmo})
\begin{eqnarray}\label{norm-cov}
\parallel\!\Psi^{(cov)}\!\!\parallel^{\,2} &=&\sum_n[m^2+\mathrm{H}(2n+1)]^{-1}\int d^6 f d^4 p\,\,\\
&&
\qquad\cdot\,\delta(f{\cdot}f^\ast)\,\delta(f{\cdot}f-2\mathrm{H}^2)\,
\delta[(p{\cdot}f^\ast)^2-\mathrm{H}^2(m^2+\mathrm{H}(2n+1))] \left|\Psi^{(cov)}(p,f) \right|^{2}\,,\nonumber
\end{eqnarray}
where the normalization constants $N_n=[m^2+\mathrm{H}(2n+1)]^{-1}$ in  (\ref{norm-cov})
are introduced in a way which leads to the independence of norm from the number $n$ describing Landau levels.

\setcounter{equation}{0}
\section{Final remarks}

\quad \, It is well known that some quantum-mechanical and field-theoretic models can be obtained by first and second quantization of
classical mechanical system. The novelty of this paper is the presentation of Landau problem describing quantum-mechanical
orbits of the particle interacting with constant magnetic field as first-quantized free motion of a particle in generalized space-time,
with Maxwell symmetries. The theory is four-dimensional and covariant by construction, and we demonstrate how to extract
from Maxwell-symmetric formulation the planar dynamics of Landau problem. In such a way we confirm
the EM origin of tensorial coordinates in Maxwell space-time and interprete the extension of Poincare symmetries to Maxwell symmetries.

We point out that in our approach there is employed new concept of enlarged space-time, with
additional dimensions linked with the description of interactions.
Maxwell algebra and additional tensorial coordinates correspond to coupling of
constant EM external field strength, but there were as well considered further
extensions of the Poincare algebra, with additional higher rank tensorial generators,
which correspond e.g. to the dipole and quadruple EM background field interactions \cite{BoGo}.
The new variety of enlarged space-times with nonvanishing torsion (or curvature) are described
by cosets of the group manifolds, generated by extended Poincare algebras.
Following Wigner successful group-theoretical definition of free elementary particles \cite{Wig}
in term of the representations of Poincare group, one can employ the enlargements of
$D=4$ Poincare symmetries with their irreducible representations as providing the
multidimensional free particles which describe equivalently in gauge-fixed frames $D=4$ particles
interacting with various classes of background fields.

In this paper we considered only spin-zero Maxwell-covariant particle.
The particles with nonvanishing spins will be considered as the next task.
One possibility consists in Maxwell-invariant generalization
of spinning particle in pseudoclassical approach with vector-like Grassmann variables (see e.g. \cite{Cas,Brink,BerMar,GerTk}).
For example, in the case of spin one-half particle we propose the following generalization of the action (\ref{act-EM-nonconst1})
\begin{eqnarray}\label{act-EM-nonconst12}
S_1 &=& \int\Big(\pi_\mu e^\mu +f_{\mu\nu}\,\omega^{\mu\nu}\Big) \\
\nonumber
&&+\,{\textstyle\frac12}\int d\tau \Big[g_1\Big(\pi{\cdot}\pi+{\textstyle\frac{i}{2}}\,e\, f_{\mu\nu}\psi^{\,\mu}\psi^{\,\nu}-m^2\Big)+ g_2\Big(f{\cdot}f-2c_2\Big)+g_3\Big(f{\cdot}f^\ast-2c_3\Big)\Big]\\
\nonumber
&&+\int d\tau \Big[{\textstyle\frac{i}{4}}\,\psi_{\mu}\dot\psi^{\,\mu} -{\textstyle\frac{i}{4}}\,\psi_{5}\dot\psi_{5}+
\chi\Big(\pi_{\mu}\psi^{\,\mu} -m\psi_{5}\Big)\Big]\,,
\end{eqnarray}
where $\psi_{\mu}$, $\psi_{5}$ and $\chi$ are Grassmann variables.
The model  (\ref{act-EM-nonconst12}) produces the constraints (\ref{phi2}), (\ref{phi3})
and modifies constraint (\ref{phi1}) as follows
\begin{equation}\label{phi1-2}
\pi{\cdot}\pi+{\textstyle\frac{i}{2}}\,e\, f_{\mu\nu}\psi^{\,\mu}\psi^{\,\nu}-m^2\approx 0\,.
\end{equation}
Besides we get the additional fermionic constraint representing pseudoclassical counterpart of Dirac eqiuation
\begin{equation}\label{ferm1-2}
\pi_{\mu}\psi^{\,\mu} -m\psi_{5}\approx 0\,.
\end{equation}
After first quantization the variables $\psi_{\mu}$, $\psi_{5}$ are realized by $4\times 4$ Dirac matrices $\gamma_\mu$ as follows:
$\hat\psi_{\mu}=\gamma_\mu \gamma_5$, $\hat\psi_{5}=\gamma_5=\gamma_0\gamma_1\gamma_2\gamma_3$.
Subsequently, the wave function is a four-component Dirac spinor $\Psi_\alpha(p,f)$ satisfying the equations
(\ref{2eq}), (\ref{3eq}) and modified equation (\ref{1eq}) which describes the Maxwell extension of
Feynman-GellMann equation \cite{FeynGel} ($\sigma^{\,\mu\nu}={\textstyle\frac{i}{2}}\,[\gamma^{\,\mu},\gamma^{\,\nu}]$)
\begin{equation}\label{1eq-12}
\Big(\pi_\mu\pi^\mu+e\, f_{\mu\nu}\sigma^{\,\mu\nu}\Big)\, \Psi=m^2\,\Psi\,.
\end{equation}
Besides, from (\ref{ferm1-2}) follows the Dirac equation
in the presence of constant EM field
\begin{equation}\label{1eq-12a}
\Big(\pi_{\mu}\gamma^{\,\mu} -m\Big)\, \Psi=0\,,
\end{equation}
where $\pi_{\mu}$ is defined by the formula (\ref{pi-q}).

It should be also noted that following \cite{GerTk} one can describe arbitrary spin of the particle equal to $N/2$
($N=1,2,\ldots$) by introducing $N$ copies
of vectorial Grassmann variables in the model. It is interesting to see whether
the first-quantized version of our model generalized to nonvanishing spins can help to
clarify the problems of interactions of higher spins with a constant EM field.

Other possible extensions of our results consists in considering the solutions of the model (\ref{act-EM-nonconst})
for the sectors described by relations (\ref{case-2}b,c),
with constant electric or radiation fields, as well as with nonvanishing Casimir $C_2$ (see (\ref{case-1})).
One can look as well for the modification of the model (\ref{act-EM-nonconst}) which would covariantize the Landau problem
with spin couplings (see e.g. \cite{Rash,Dress}).

It is interesting to study whether the Maxwell enlargement of Poincare symmetries has also other applications.
The presence of additional tensorial coordinates by analogy with
\cite{BL,BLS,Vas,PST} suggests their
possible use to the description of higher spin fields.
First step in such a direction was made by present authors in recent paper \cite{FedLuk}.
In this paper we consider the spinorial particle model in Maxwell space and obtain after
first quantization the infinite expansion of the wave function into
the interacting higher spin fields.
Another application of the Maxwell symmetries with possible cosmological implications (see \cite{AKLuk,SS-D,Durka,Durka2})
is the extension of standard gauge approach to gravity by supplementing the Einstein-Cartan gravity framework with additional geometric
Abelian gauge fields obtained by gauging the six symmetries generated by generators $Z_{\mu\nu}$.
It is a matter of further studies to find the possible role of these new vector fields in cosmology and astrophysics.

\section*{Acknowledgements}

\noindent
The authors would like to thank Evgeny Ivanov and Piotr Kosi\'{n}ski for valuable remarks, as well as Dmitri Sorokin for his interest in our work
and long correspondence on possible link between particle models in Maxwell tensorial space and
higher spin fields.
We acknowledge a support from the grant of the Bogoliubov-Infeld Programme and
RFBR grants 09-01-93107, 11-02-90445, 12-02-00517 (S.F.), as well as
from the Polish Ministry of Science and Higher Educations grant No.~N202331139 (J.L.).
S.F. thanks the members of the Institute of Theoretical Physics at Wroclaw University
for their warm hospitality.

\end{document}